\documentclass[vecarrow]{svmult}

\usepackage[utf8]{inputenc}

\usepackage{mathptmx}       % selects Times Roman as basic font
\usepackage{helvet}         % selects Helvetica as sans-serif font
\usepackage{courier}        % selects Courier as typewriter font
\usepackage{amsmath,amssymb}
\usepackage{makeidx}         % allows index generation
\usepackage{graphicx}        % standard LaTeX graphics tool
\usepackage{multicol}        % used for the two-column index
\usepackage[bottom]{footmisc}% places footnotes at page bottom

\makeindex             % used for the subject index
\newcommand{\eg}{e.g.}
\newcommand{\ie}{i.e.}
\newcommand{\Ls}{L_{\text{s}}}
\newcommand{\Lth}{L_{\text{th}}}
\newcommand{\Zh}{z_{\text{h}}}

\begin{document}

\title*{Heavy Probes in Strongly Coupled Plasmas\\With Chemical Potential}
\author{Carlo Ewerz and Andreas Samberg}
\institute{%
  Carlo Ewerz, \email{c.ewerz@thphys.uni-heidelberg.de}\\%
  Andreas Samberg, \email{a.samberg@thphys.uni-heidelberg.de}\\
  \at Institut f\"ur Theoretische Physik, Universit\"at Heidelberg, Philosophenweg 16,\\D-69120 Heidelberg, Germany
  \at ExtreMe Matter Institute EMMI, GSI Helmholtzzentrum f\"ur Schwerionenforschung,\\Planckstra{\ss}e 1, D-64291 Darmstadt, Germany
}
\maketitle

\abstract*{We study the properties of heavy probes moving in strongly
  coupled plasmas at finite chemical potential. Using the
  gauge/gravity duality we consider large classes of gravity models
  consisting in deformed AdS$_5$ spacetimes endowed with
  Reissner--Nordström-type black holes.  We report on our analysis of
  the screening distance of a quark--antiquark pair, its free energy,
  and the running coupling.
  These observables show a certain insensitivity as to which model and
  deformation is used, pointing to strong-coupling universal behavior.
  Thus, the results may be relevant for modeling heavy quarkonia
  traversing a quark--gluon plasma at finite net baryon density, and
  their suppression by melting.%
}

\abstract{We study the properties of heavy probes moving in strongly
  coupled plasmas at finite chemical potential. Using the
  gauge/gravity duality we consider large classes of gravity models
  consisting in deformed AdS$_5$ spacetimes endowed with
  Reissner--Nordström-type black holes.  We report on our analysis of
  the screening distance of a quark--antiquark pair, its free energy,
  and the running coupling.
  These observables show a certain insensitivity as to which model and
  deformation is used, pointing to strong-coupling universal behavior.
  Thus, the results may be relevant for modeling heavy quarkonia
  traversing a quark--gluon plasma at finite net baryon density, and
  their suppression by melting.%
}

%%%%%%%%%%%%%%%%%%%%%%%%%%%%%%%%%%%%%%%%%%%%%%%%%%%%%%%

\section{Introduction}
\label{sec:introduction}

Over the past years, gauge/gravity duality\index{gauge-gravity duality}
(\cite{Maldacena:1997re,Gubser:1998bc,Witten:1998qj}; see
\eg~\cite{CasalderreySolana:2011us} for a review) has been
successfully applied to the physics of the QCD medium created in heavy
ion collisions at RHIC and LHC. One of the most prominent theoretical
results is the computation of a universal value for the ratio of the
shear viscosity $\eta$ to the entropy density $s$ in a large class of
strongly coupled deconfined plasmas
\cite{Policastro:2001yc,Kovtun:2003wp} and the conjecture that this
value ($\eta/s = 1/(4\pi)$) may be a lower bound for all physical
substances \cite{Kovtun:2004de}. On the experimental side, it was
found that the Quark--Gluon Plasma (QGP) created at RHIC and LHC has a
value of $\eta/s$ in the ballpark of the holographic result,
indicating strong coupling at the accessible temperatures.

Beyond current high-energy heavy ion programs, which mainly study the
high-temperature, low-chemical potential regime of QCD, a major open
question is the phase structure of QCD at nonzero baryon chemical
potential\index{chemical potential}. Experimentally, this will be
addressed for example at future FAIR experiments. %
Theoretically, the ability to handle nonzero chemical potential in QCD
or at least QCD-like theories is crucial. %

We explore, by means of the gauge/gravity duality, the phase structure
of strongly coupled non-conformal theories similar to QCD by
investigating the physics of probes in the thermal plasmas of these
theories with nonzero chemical potential. In a spirit similar to the
holographic computation of $\eta/s$ we search for
universality\index{universality} in the behavior of heavy
quark--antiquark ($Q\bar Q$) bound states in large classes of
holographic theories. This may yield insight into certain QCD
processes relevant for the QGP produced in heavy ion collisions, \eg\
suppression of charmonia \cite{Matsui:1986dk} or bottomonia.

Specifically, we analyze the $Q\bar Q$ screening
distance\index{screening distance}, the $Q\bar Q$ free energy in the
medium (roughly speaking the interaction potential\index{interaction
  potential}), and the running coupling\index{running coupling}
extracted from the free energy.
Previous work on similar problems includes
\cite{Liu:2006nn,Liu:2008tz} for vanishing chemical potential, and
\cite{Caceres:2006ta,Avramis:2006em} for nonzero chemical potential in
$\mathcal{N}=4$ supersymmetric Yang--Mills theory.
A more detailed account of our findings will be published elsewhere.

We start with the prototype of gauge/gravity
duality\index{gauge-gravity duality} between classical supergravity on
AdS$_5$ and conformal $\mathcal{N}=4$ supersymmetric Yang--Mills
theory (SYM) with gauge group SU$(N_{\text{c}})$ in the limit of
infinite number of colors, $N_{\text{c}}\rightarrow\infty$, and large
't Hooft coupling $\lambda \equiv g_{\text{YM}}^2 N_{\text{c}}$. A
thermal bath for the gauge theory is dual to a black hole in
AdS$_5$. Putting charge on the black hole, we can induce a chemical
potential in the dual theory. Therefore, as a starting point we
consider AdS$_5$ with a Reissner--Nordström black hole (AdS-RN), in
Poincar\'e coordinates,
\begin{align}
  \D s^{2} &= \frac{R^2}{z^2} \left( -h(z)\D t^{2} + \D \vec{x}^2 + \frac{\D z^2}{h(z)} \right) \label{Intro:N=4metric}\\
  \text{with}\quad h(z) &= 1 - \left( 1 + Q^2\right) \left(\frac{z}{\Zh}\right)^4 + Q^2 \left(\frac{z}{\Zh}\right)^6 \,.\nonumber
\end{align}
Here, $R$ is the curvature scale and the black hole horizon is at
$z=\Zh$. The dual theory is in a thermal state with temperature $T =
(1-\frac{1}{2}Q^2)/(\pi\Zh)$ and chemical potential $\mu =
\sqrt{3}Q/\Zh$. We have $0 \le Q \le \sqrt{2}$.

To come closer to real-world physics, we study models in which
conformality is explicitly broken by deforming the AdS spacetime. On
the one hand, we consider the \emph{CGN model} proposed by Colangelo,
Giannuzzi, and Nicotri \cite{Colangelo:2010ho}. It is specified by the
metric \eqref{Intro:N=4metric} with an additional overall warp factor
$\E^{c^2z^2}$ with deformation parameter $c$. On the other hand, we
study a family of \emph{1-parameter models} which we derive from the
action used in \cite{DeWolfe:2010he}, which adds to 5-dimensional
gravity with metric $g_{\mu\nu}$ and negative cosmological constant
a scalar field $\phi$ and a U$(1)$ gauge field $A_\mu$ whose boundary
value equals the chemical potential\index{chemical potential} in the
dual gauge theory. Our ansatz with deformation parameter $\kappa$ is
\begin{align}
  \label{eq:4}
  g_{\mu\nu}\D x^\mu \D x^\nu = \E^{2 A(z)} \left( -h(z)\D t^2 + \D\vec{x}^2 \right) + \frac{\E^{2 B(z)}}{h(z)} \D z^2\,,\\
  A(z) = \log\left(\frac{R}{z}\right),\qquad \phi(z) = \sqrt{\frac{3}{2}} \kappa z^2,\qquad A_\mu\D x^\mu = \Phi(z)\D t\,,
\end{align}
where $R$ is a constant and $h(z)$ is the redshift factor induced by
the black hole. 
We derive two classes of models from this ansatz, treating $\phi$ as
the dilaton or not, called `string frame' and `Einstein frame' models,
respectively. The solutions for $B, h$, and $\Phi$ can be given in
closed form \cite{Samberg:2012da}. %
At fixed $(\mu, T)$, a maximal deformation $\kappa_{\text{max}}$
exists that still allows a black hole solution representing $(\mu,
T)$. %
Since these models solve gravity equations of motion (EOMs) they are
expected to be thermodynamically consistent, as opposed to models in
which the metric is deformed `by hand'.

In the CGN model we find unusual behavior in some observables at low
temperatures and chemical potentials. Such behavior occurs when we
consider moving probes, for example in the case of the drag force. We
believe that these artifacts are unphysical. Indeed, they do no longer
occur in models obtained as solutions of gravity EOMs. In particular,
they are absent in our 1-parameter models. We will report on this in
detail elsewhere \cite{ELS13}.

%%%%%%%%%%%%%%%%%%%%%%%%%%%%%%%%%%%%%%%%%%%%%%%%%%%%%%%

\section{Screening Distance}
\label{sec:screening-distance}

We study a dipole of an infinitely heavy quark and its antiquark,
separated by a distance $L$ in the deconfined plasma of the gauge
theory. The quarks are situated at the 4-dimensional boundary ($z=0$
in our coordinates) and are connected by a macroscopic string in the
bulk (see \eg~\cite{CasalderreySolana:2011us}). %
We accommodate a finite velocity $v$ of the $Q\bar Q$ system with
respect to the surrounding medium by boosting the bulk metric with
rapidity $\eta = \operatorname{artanh}(v)$. %
In order to find the classical string configuration we have to
extremize the Nambu--Goto action in the given gravity background.

There is a distance $\Ls$, such that for $L < \Ls$ there are two
string configurations connecting the dipole, while no such solution
exists for $L > \Ls$. Thus, $\Ls$ is called the \emph{screening
  distance}\index{screening distance} of the $Q\bar Q$ interaction in
the thermal medium.  In the studies
\cite{Schade:2012ah,SchadeEwerzConfX} it was found that at any
temperature $T$ the screening distance is bounded from below by
$\Ls^{\mathcal{N}=4\text{ SYM}}(T)$ under consistent deformations of
AdS-Schwarzschild, the dual of $\mathcal{N}=4$ SYM at $\mu=0$ (where
we understand consistency in the sense of solving equations of motion
of a suitable 5-dimensional gravity action).

The question arises whether this bound holds under the inclusion of a
chemical potential. We find that in the CGN and similar models,
depending on (sign) choices in the metric, $\Ls$ can change in both
directions. More interesting are consistent deformations which in our
case are the two classes of 1-parameter models.

\begin{figure}[t]
%  \centering
  \sidecaption[t]
  \includegraphics[width=7.5cm]{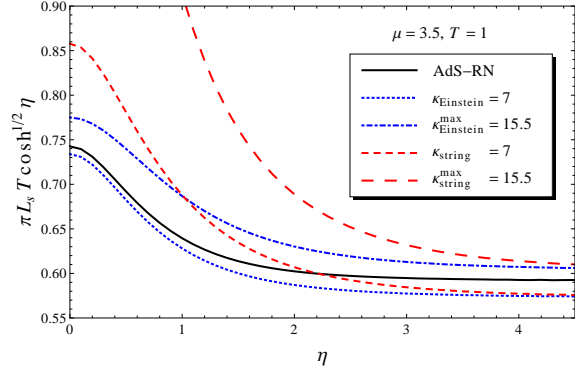}
  \caption{%
    Rapidity dependence of the screening distance $\Ls$ at finite
    temperature $T$ and chemical potential $\mu$ in the 1-parameter
    models, evolving from the conformal case (dual to AdS-RN) to the
    maximal deformation $\kappa_{\text{max}}$. The asymptotic behavior
    $\Ls \propto \cosh^{-1/2}(\eta)$ is scaled out. At large $\eta$,
    the differences between the two 1-parameter models vanish. $\Ls$,
    $\mu$, and $\kappa$ are measured in units of temperature.%
  }
  \label{fig:screeningDist1p}
\end{figure}

As can be seen from Fig.~\ref{fig:screeningDist1p}, the bound on $\Ls$
is violated in the Einstein frame models at large chemical potential,
approximately when $\mu \gtrsim \sqrt\kappa$. In the string frame
models it is only violated for $Q\bar Q$ pairs moving sufficiently
fast at large $\mu \gtrsim \sqrt\kappa$.

However, for all deformations, the amount of violation of the bound is
relatively small such that there might exist a slightly lower,
improved bound. Moreover, in the Einstein frame models, the screening
distance is a robust observable, depending only weakly on the
deformation.

We also study the dependence of $\Ls$ on the $Q\bar Q$ rapidity
$\eta$. Figure \ref{fig:screeningDist1p} illustrates that the
ultrarelativistic scaling of the screening distance $\Ls \propto
\cosh^{-1/2}(\eta)$ is robust and remains valid in all models, at all
chemical potentials. It is interesting to note that this robustness
against deformation is different from what was found in
\cite{Caceres:2006ta} for other explicitly non-conformal models.

%%%%%%%%%%%%%%%%%%%%%%%%%%%%%%%%%%%%%%%%%%%%%%%%%%%%%%%

\section{Free Energy and Running Coupling}
\label{sec:free-energy}

\begin{figure}[t]
%  \centering
  \sidecaption[t]
  \includegraphics[width=7.5cm]{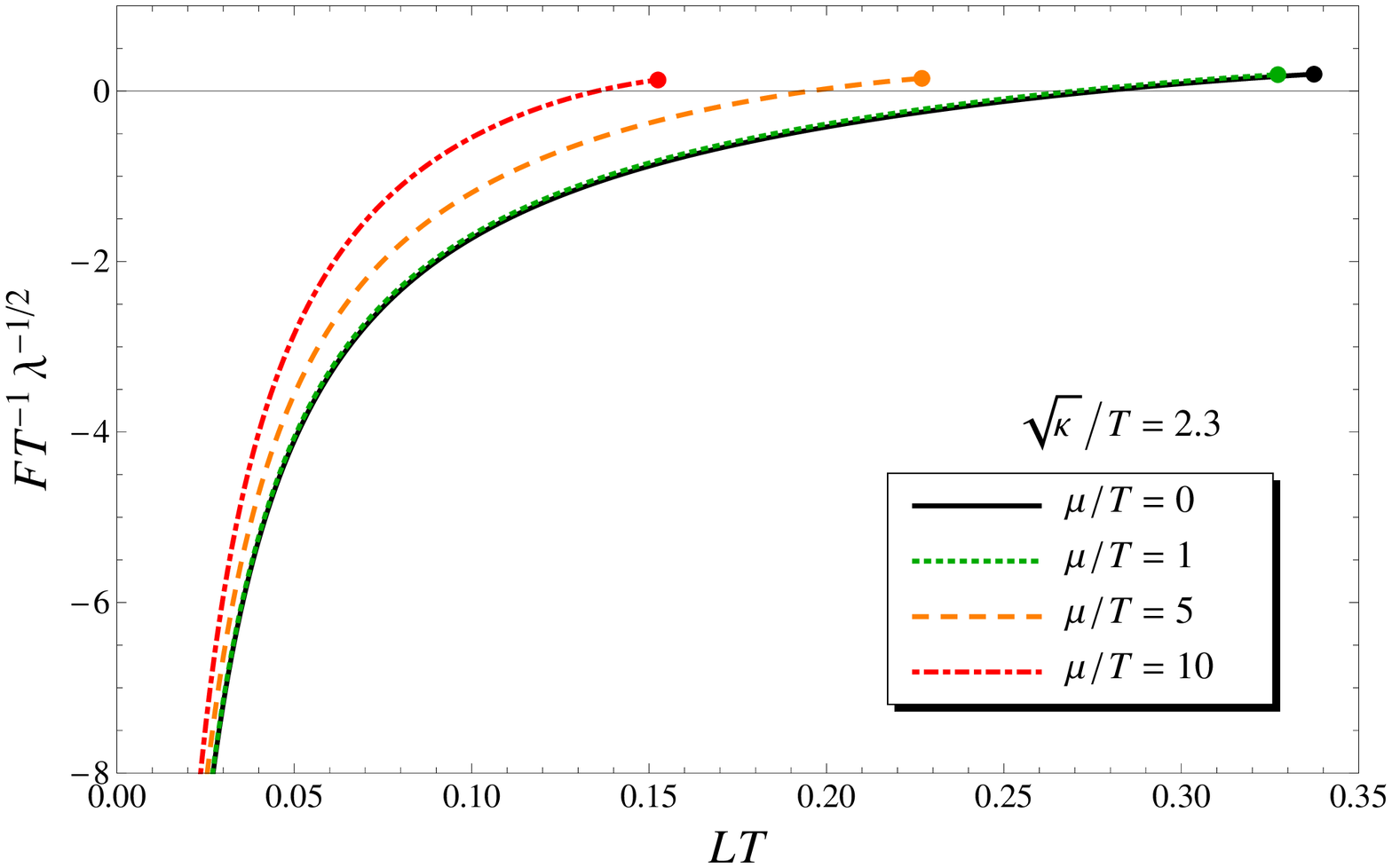}
  \caption{%
    Typical $Q\bar Q$ free energy $F(L)$ in a non-conformal
    1-parameter string frame model for varying values of the chemical
    potential $\mu$. The behavior in the Einstein frame models is
    qualitatively similar. $\sqrt\lambda$ is a constant. The endpoints
    of the curves are located at the respective screening distances.%
  }
  \label{fig:freeEn1pStr}
\end{figure}

The free energy $F(L)$\index{free energy} of the $Q\bar Q$ system can
be extracted from the extremal classical string action following a
well-known procedure (see \eg~\cite{CasalderreySolana:2011us}). The
typical features in the consistently deformed models can be seen in
Fig.~\ref{fig:freeEn1pStr}. Here, $F(L)$ is normalized such that a
configuration having $F < 0$ has less free energy than the
non-interacting, unbound $Q\bar Q$ system. Thus, we see that the
effect of increasing the chemical potential is a decrease in binding
energy.
Taking this together with our findings concerning the screening
distance, we see that in these holographic models an increased net
density around the $Q\bar Q$ dipole weakens its binding by screening
the interaction. We find this effect regardless of the specific model
under consideration.

To explore the impact of non-conformality on the interaction in more
detail, we study the running coupling\index{running coupling} defined
via the derivative of the free energy,
\begin{equation}
  \label{eq:1}
  \alpha_{Q\bar Q}(L) \equiv \frac{3}{4} L^2\frac{\D F(L)}{\D L} \,.
\end{equation}
In the conformal case, where $F(L) \propto 1/L$ is Coulombic,
$\alpha_{Q\bar Q} = \text{const}$. Hence, any non-trivial dependence
on $L$ measures the deviation from conformality.

\begin{figure}[t]
  \centering
  \includegraphics[width=.98\linewidth]{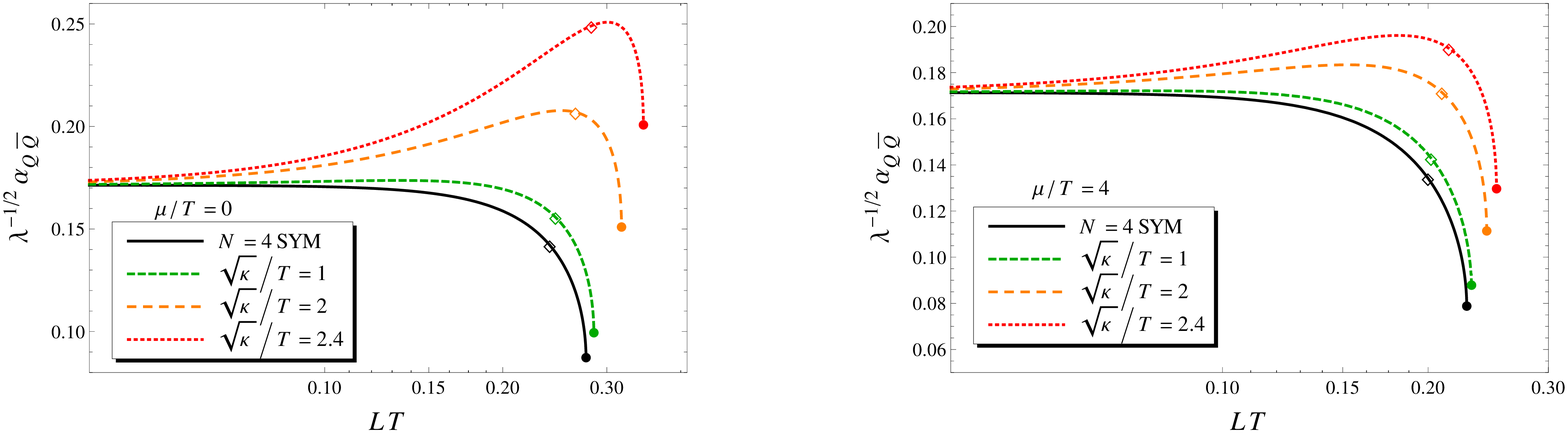}
  \caption{%
    Running coupling $\alpha_{Q\bar Q}(L)$ in $\mathcal{N}=4$ SYM and
    1-parameter string frame models with increasing non-conformality
    parameter $\kappa$, for zero (left) and nonzero (right) chemical
    potential $\mu$.%
  }
  \label{fig:runncoup}
\end{figure}

In Fig.~\ref{fig:runncoup} we show $\alpha_{Q\bar Q}(L)$ for the
1-parameter string frame models. The qualitative picture is the same
in the Einstein frame models; however, in these models, $\alpha_{Q\bar
  Q}$ is very robust under deformations, so that the curves for
different $\kappa$ deviate only very little from each other.

From Fig.~\ref{fig:runncoup} we see that $\alpha_{Q\bar Q}$ is
constant in the UV, \ie~at small distances $L$. One can also see this
from the free energy itself in Fig.~\ref{fig:freeEn1pStr} which
approaches Coulombic form in the UV. This is due to the restoration of
conformality in the UV, the bulk realization of which is the condition
on the metric to be asymptotically AdS.

At larger distance, both the explicit non-conformality and the thermal
medium affect the $Q\bar Q$ interaction. In effect, the running
coupling starts to deviate from its asymptotic value.
We find a robust increase above the UV value at intermediate length
scales due to non-conformality in all deformed models, at vanishing
$\mu$ \cite{Schade:2012ah} but also at nonzero chemical potential.

The plasma starts to take effect at the thermal scale, roughly
$\Lth\sim 1/T$, leading to a drop-off of $\alpha_{Q\bar Q}$, before
the $Q\bar Q$ interaction is entirely screened. (The endpoint of the
curves $\alpha_{Q\bar Q}(L)$ is the screening distance $\Ls$.)
This pattern is also found in lattice QCD studies of $\alpha_{Q\bar
  Q}$ in the deconfined phase \cite{Kaczmarek:2004gv}.
While the lattice calculations are presently restricted to vanishing
chemical potential due to the sign problem, the holographic models
allow us to explore the effect of a chemical potential in strongly
coupled QCD-like theories.

We find that the effect of the chemical potential is relatively weak:
the drop-off scale is only weakly dependent on the chemical potential
while it strongly depends on the temperature.

%%%%%%%%%%%%%%%%%%%%%%%%%%%%%%%%%%%%%%%%%%%%%%%%%%%%%%%

\section{Conclusion}
\label{sec:conclusion}

We have reported on some results of our studies of deformed, \ie\
non-conformal, gauge/\allowbreak{}gravity models for strongly coupled
plasmas at nonzero chemical potential. Studying large classes of
holographic models, we address the problem of the strong coupling
dynamics of moving heavy mesons in deconfined plasmas by looking for
universality. In particular, we include a nonzero chemical potential
$\mu$ in our studies.

We find a certain robustness of the screening distance $\Ls$ at
nonzero $\mu$ under deformations\index{universality}. However, when
switching on the chemical potential, $\Ls^{\mathcal{N}=4\text{ SYM}}$
no longer is a lower bound on the screening distance under
deformations, unlike in the case of $\mu=0$.

Furthermore, we observe a weak impact of the chemical potential on the
qualitative features of the quark--antiquark interaction. Even
quantitatively, the dependence of characteristic scales on the
chemical potential is generally significantly weaker than their
dependence on temperature.

%%%%%%%%%%%%%%%%%%%%%%%%%%%%%%%%%%%%%%%%%%%%%%%%%%%%%%%

\begin{acknowledgement}
  We thank K.~Schade for many helpful discussions. This work was
  supported by the ExtreMe Matter Institute EMMI.
\end{acknowledgement}

%%%%%%%%%%%%%%%%%%%%%%%%%%%%%%%%%%%%%%%%%%%%%%%%%%%%%%%

% \bibliographystyle{spphys}
% \bibliography{lit_red}

\end{document}